\documentclass[fleqn,12pt,twoside]{article}
\usepackage[headings]{espcrc1}

\readRCS
$Id: espcrc1.tex,v 1.2 2004/02/24 11:22:11 spepping Exp $
\usepackage{graphicx}
\usepackage[figuresright]{rotating}

\newcommand{\AmS}{{\protect\the\textfont2
  A\kern-.1667em\lower.5ex\hbox{M}\kern-.125emS}}

\title{The QCD equation of state for two flavours 
at non-zero chemical potential}

\author{S.~Ejiri\address[TKY]{Department of Physics, 
The University of Tokyo, Tokyo 113-0033, Japan}%
\thanks{Speaker. \hspace{1mm} SE is supported by Grants-in-Aid of the Japanese 
MEXT No.~15540254. KR is supported by KBN under grant 2P03 (06925). 
FK is partly supported by a contract DE-AC02-98CH1-886 with the U.S. DOE.
MD is supported by DFG grant GRK-881.
},
C.R.~Allton\address[SWA]{Department of Physics, University of Wales Swansea,
Swansea SA2 8PP, UK},
M.~D\"{o}ring\address[BIE]{Fakult\"at f\"ur Physik, Universit\"at 
Bielefeld, D-33615 Bielefeld, Germany}\address[BNL]{Physics Department, 
Brookhaven National Laboratory, Upton, NY 11973, USA},
S.J.~Hands\addressmark[SWA], O.~Kaczmarek\addressmark[BIE],
F.~Karsch\addressmark[BIE]\addressmark[BNL], E.~Laermann\addressmark[BIE] and 
K.~Redlich\address[WRO]{Institute of Theoretical Physics, 
University of Wroclaw, PL-50204 Wroclaw, Poland}\address[CERN]{Physics 
Department, Theory Division, CERN, CH-1211 Geneva 23, Switzerland}.
}

\runtitle{The QCD equation of state for two flavours 
at non-zero chemical potential}
\runauthor{S. Ejiri et al.}

\begin{document}

\maketitle

\begin{abstract}
We present results of a simulation of 2 flavour QCD on a $16^3\times4$
lattice using p4-improved staggered fermions with bare quark mass 
$m/T=0.4$.  Derivatives of the thermodynamic grand canonical partition
function  $Z(V,T,\mu_u,\mu_d)$ with respect to chemical potentials 
$\mu_{u,d}$ for different quark flavours are calculated up to sixth order, 
enabling estimates of the pressure and the quark number density as well 
as the chiral condensate and various susceptibilities as functions 
of $\mu_{u,d}$ via Taylor series expansion. 
Results are compared to high temperature perturbation theory as well
as a hadron resonance gas model. 
We also analyze baryon as well as isospin fluctuations and discuss 
the relation to the chiral critical point in the QCD phase diagram. 
We moreover discuss the dependence of the heavy quark free energy on 
the chemical potential.
\end{abstract}

\section{Introduction}

It is important to study QCD at high temperature and high density by 
numerical simulations of lattice QCD. 
In particular, studies of the equation of state (EoS) can provide basic input 
for the analysis of the experimental signatures for QGP formation, e.g. 
the EoS will control the properties of any hydrodynamic expansion. 
We extend studies of the EoS to non-zero baryon number density.

Simulation at non-zero baryon density is known to be difficult; 
however recent studies have found that a Taylor expansion with respect 
to chemical potential $\mu_{u,d}$ is an efficient technique to 
investigate the low density regime, interesting for heavy-ion collisions 
\cite{eos1,eos2}. 
In the calculation of the Taylor expansion coefficients, i.e. calculation 
of the derivatives at $\mu_{u,d}=0$, the technical difficulty for non-zero 
$\mu_{u,d}$ does not arise and a quantitative study becomes possible within 
the error by truncation of higher order terms. 
Since thermodynamic quantities can be defined by derivatives of the 
partition function, e.g. chiral condensate 
$\langle \bar{\psi} \psi \rangle = (T/V) (\partial \ln Z/ \partial m)$, 
quark number density 
$n_{q} = (T/V) (\partial \ln Z/ \partial \mu_q)$, 
quark number susceptibility 
$\chi_{q} = (T/V) (\partial^2 \ln Z/ \partial \mu_q^2)$, and
isospin susceptibility 
$\chi_{I} = (T/V) (\partial^2 \ln Z/ \partial \mu_I^2)$, 
where $\mu_q = (\mu_u +\mu_d)/2$ and $\mu_I = (\mu_u-\mu_d)/2$,
the calculation of first derivatives yields basic QCD thermodynamics 
observables, and the calculation of higher derivatives is a natural extension. 

In this study, we evaluate the Taylor expansion coefficients of 
thermodynamic quantities and compare these results to the predictions from 
the perturbation theory in the high temperature limit and from the hadron 
resonance gas model in the low temperature hadron phase. 
Also fluctuations of quark number, isospin number and electric charge 
are discussed near the transition temperature $T_c$. 
They are estimated by $\chi_q$, $\chi_I$ and 
charge susceptibility $\chi_{C} = \chi_q/36 + \chi_I/4$, 
and are related to the critical endpoint expected at non-zero $\mu_q$. 
Moreover we study the Taylor expansion coefficients of chiral condensate 
and heavy quark free energy. 
The details of simulations are given in \cite{eos2}.

\begin{figure}[tb]
\begin{center}
\includegraphics[width=5cm]{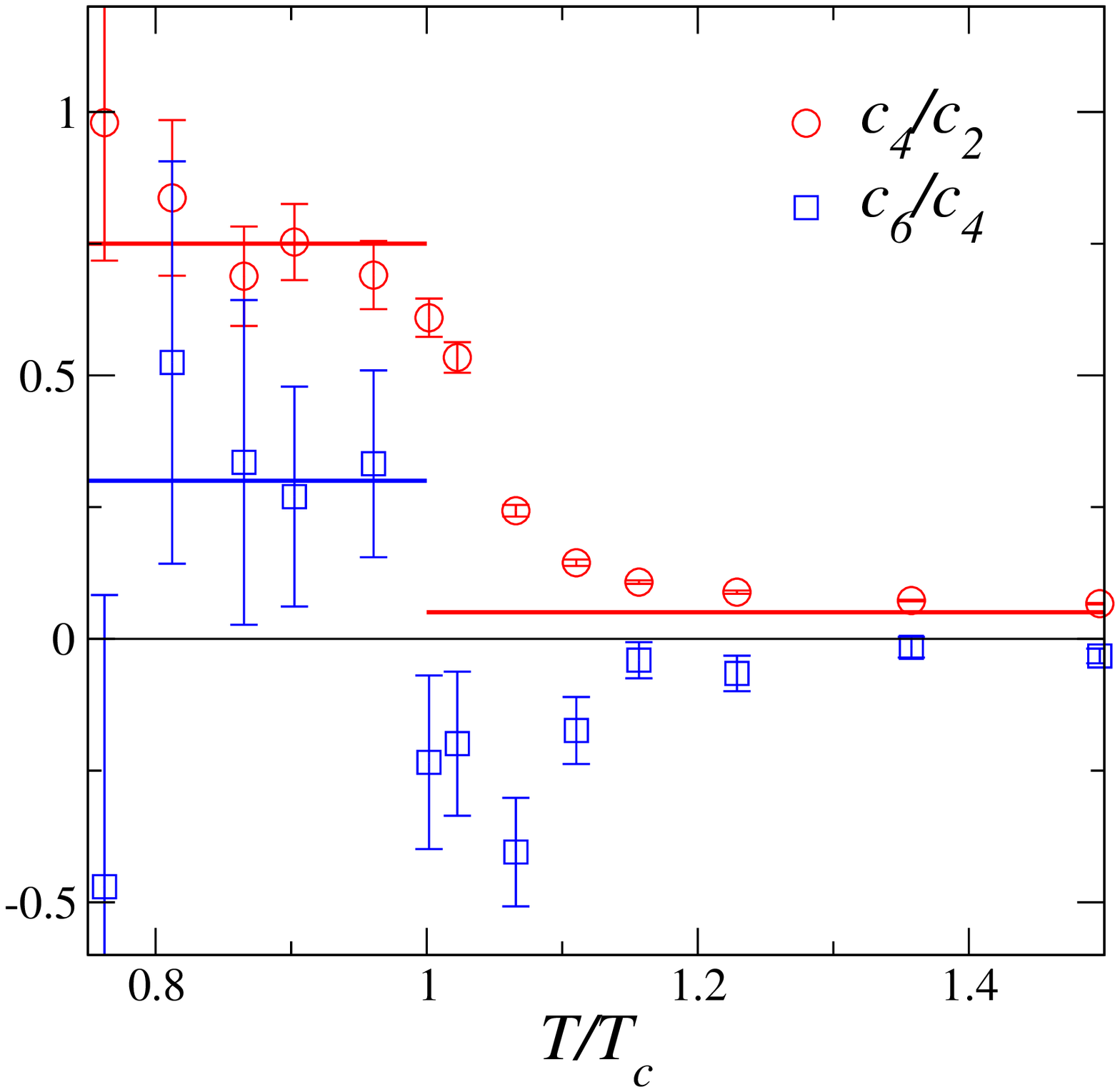}
\includegraphics[width=5cm]{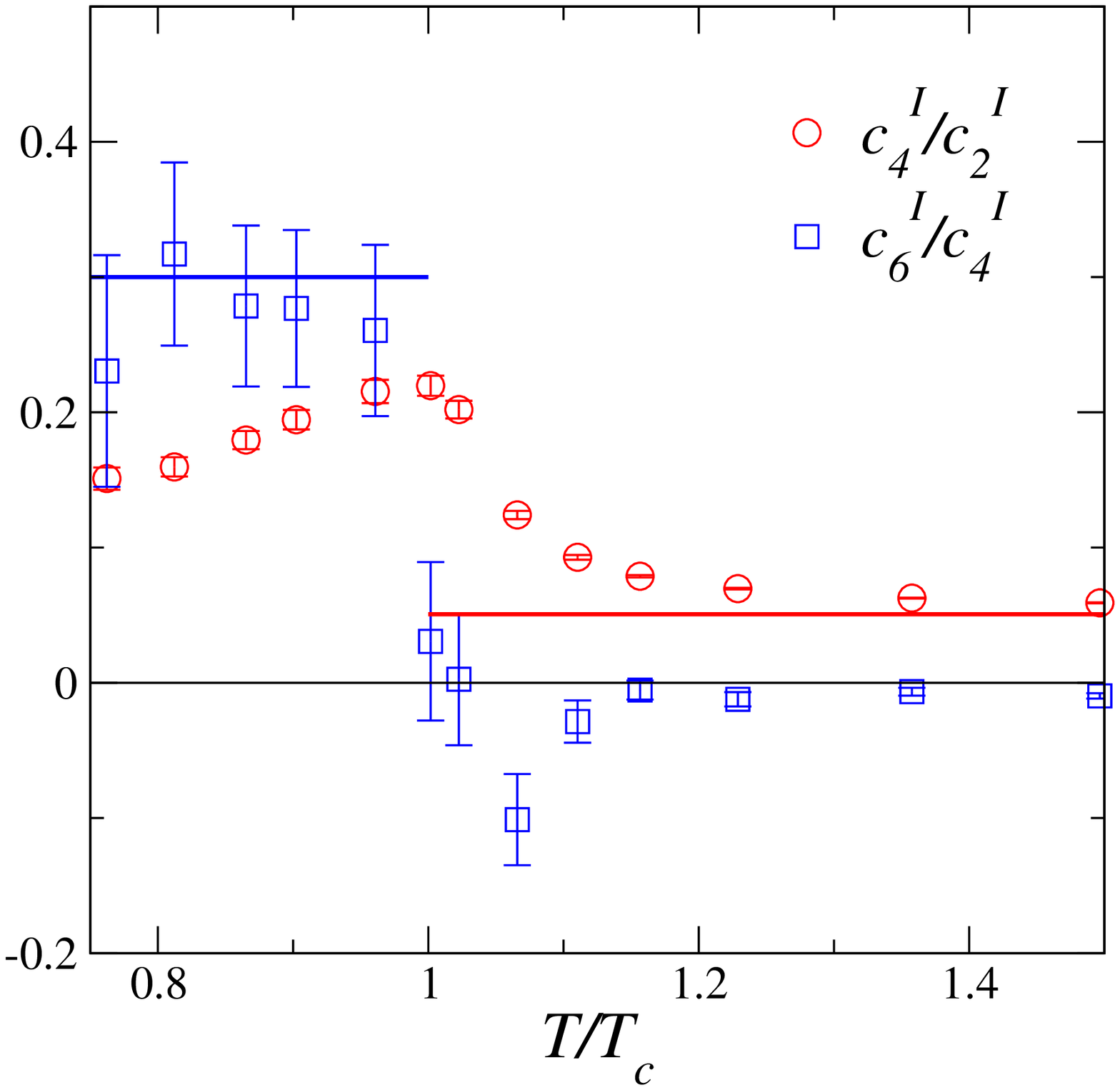}
\includegraphics[width=5cm]{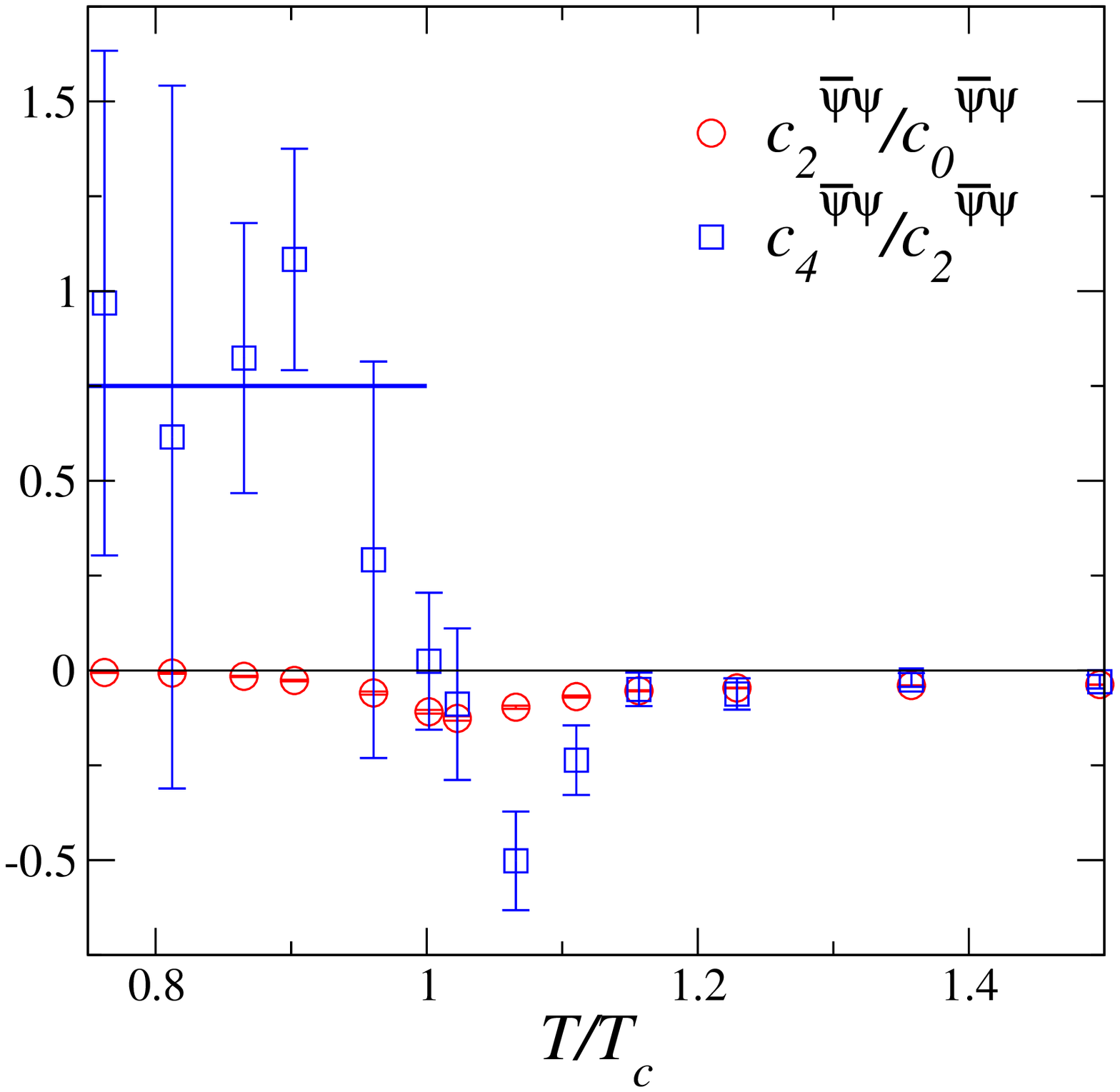}
\vskip -0.8cm
\caption{The ratios of Taylor expansion coefficients for pressure 
(quark number susceptibility), isospin susceptibility and chiral condensate.}
\end{center}
\vskip -0.6cm
\label{fig:c2c4c6}
\end{figure}

\section{Quark gluon gas and hadron resonance gas}

We define the expansion coefficients 
by $p/T^4 =(\ln Z)/(VT^3) 
\equiv \sum_{n=0}^{\infty} c_n (\mu_q/T)^n$, 
$\chi_q [\chi_I]/T^2 \equiv 
\sum_{n=2}^{\infty} n(n-1) c_n [c_n^I] (\mu_q/T)^{n-2}$ and 
$\langle \bar{\psi}\psi \rangle /T^3 
\equiv \sum_{n=0}^{\infty} c_n^{\bar{\psi}\psi} (\mu_q/T)^n$ 
for $\mu_I=0$.
We expect the equation of state to approach that of a free quark-gluon 
gas (Stefan-Boltzmann (SB) gas) in the high temperature limit. 
The coefficients in the SB limit for 2 flavour QCD with $\mu_I=0$ 
are well known as 
$c_2=c_2^I=1, c_4=c_4^I =1/(2\pi^2)$ and $c_n=c_n^I=0$ for $n \geq 6$. 
Moreover the sign of $c_6$ is negative starting at $O(g^3)$ 
in perturbation theory.
On the other hand, in the low temperature phase QCD is well described 
by a hadron resonance gas model in which 
the pressure is obtained by summing over the contributions from 
all resonance states of hadrons. In this model, the contribution 
to $p/T^4$ from each individual hadron which has baryon number $B_i$ is 
in proportion to $\exp(3B_i \mu_q /T)$, 
hence the pressure can be written as 
$p/T^4=G(T)+F(T) \cosh (3 \mu_q/T)$, where $G(T)$ and 
$F(T)$ are the mesonic and baryonic components of $p/T^4$ at $\mu_q=0$. 
Similarly, we obtain $\chi_q/T^2=9F(T) \cosh (3 \mu_q/T)$, 
$\chi_I/T^2=G^I(T)+F^I(T) \cosh (3 \mu_q/T)$ and
$\langle \bar{\psi}\psi \rangle/T^3=G^{\bar{\psi}\psi}(T)
+F^{\bar{\psi}\psi}(T) \cosh (3 \mu_q/T)$. 
Here, the mesonic component for $\chi_q$ is zero because mesons 
have baryon number zero. 
Therefore, $c_4/c_2=3/4$, $c_6/c_4=3/10$, $c_6^I/c_4^I=3/10$ and 
$c_4^{\bar{\psi}\psi}/c_2^{\bar{\psi}\psi}=3/4$ in the region 
where the hadron resonance gas provides a good approximation. 

We investigate these coefficients. 
The results for $c_{n+2}/c_n$, $c_{n+2}^I/c_n^I$ and 
$c_{n+2}^{\bar{\psi}\psi}/c_n^{\bar{\psi}\psi}$ 
are shown in Fig.~\ref{fig:c2c4c6}. 
We find that these results are consistent with the prediction from 
the hadron resonance gas model for $T/T_c \leq 0.96$ and
approach the SB values, 
i.e. $c_4/c_2= c_4^I/c_2^I=1/2\pi^2$, $c_6/c_4=c_6^I/c_4^I=0$, 
in the high temperature limit. Also $c_6$ is negative at high 
temperature as expected in the perturbation theory. 
These results suggest that the models of free quark-gluon gas
and hadron gas seem to explain the behaviour of thermodynamic 
quantities well except in the narrow regime near $T_c$.

\begin{figure}[tb]
\begin{center}
\includegraphics[width=7.0cm]{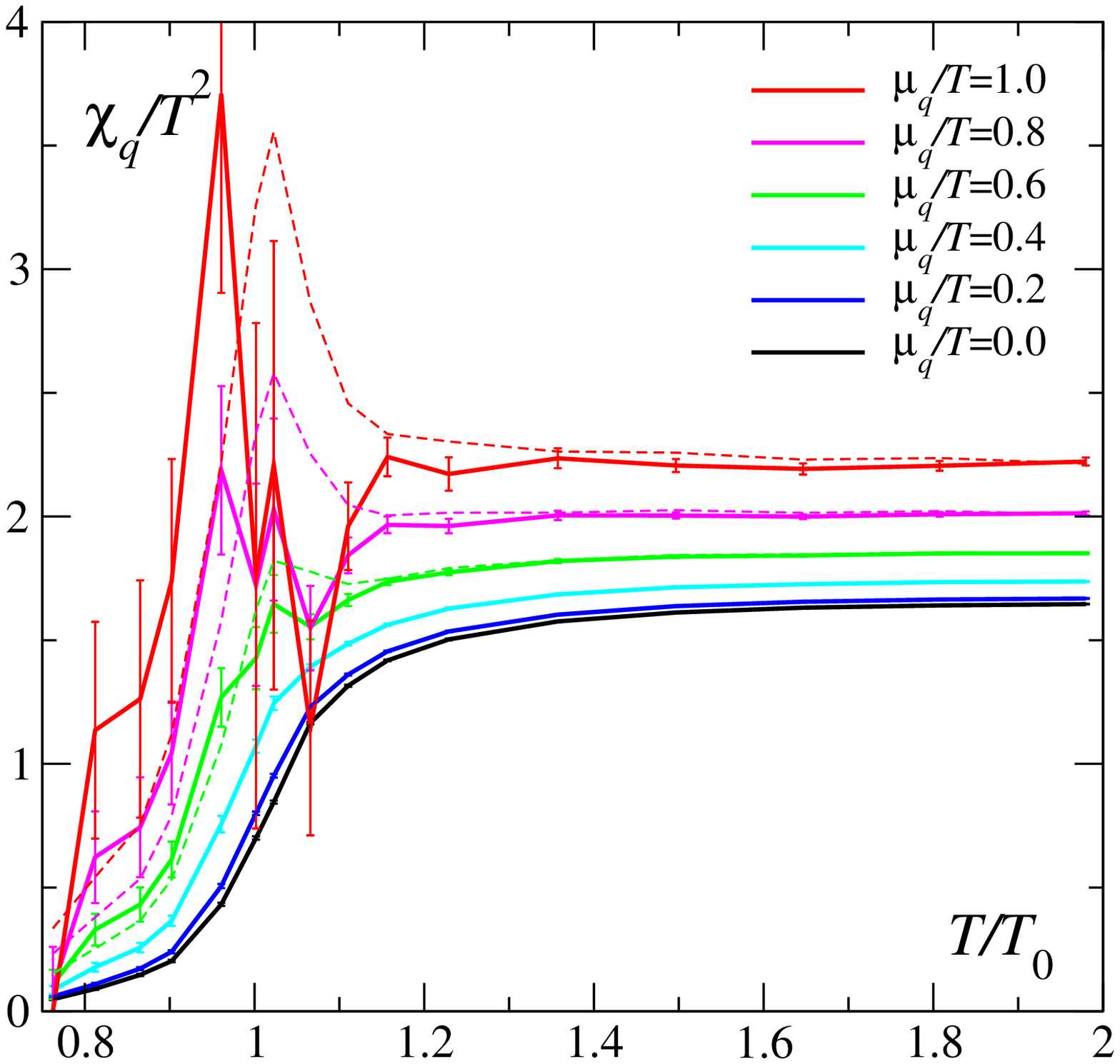}
\hskip 0.7cm
\includegraphics[width=7.0cm]{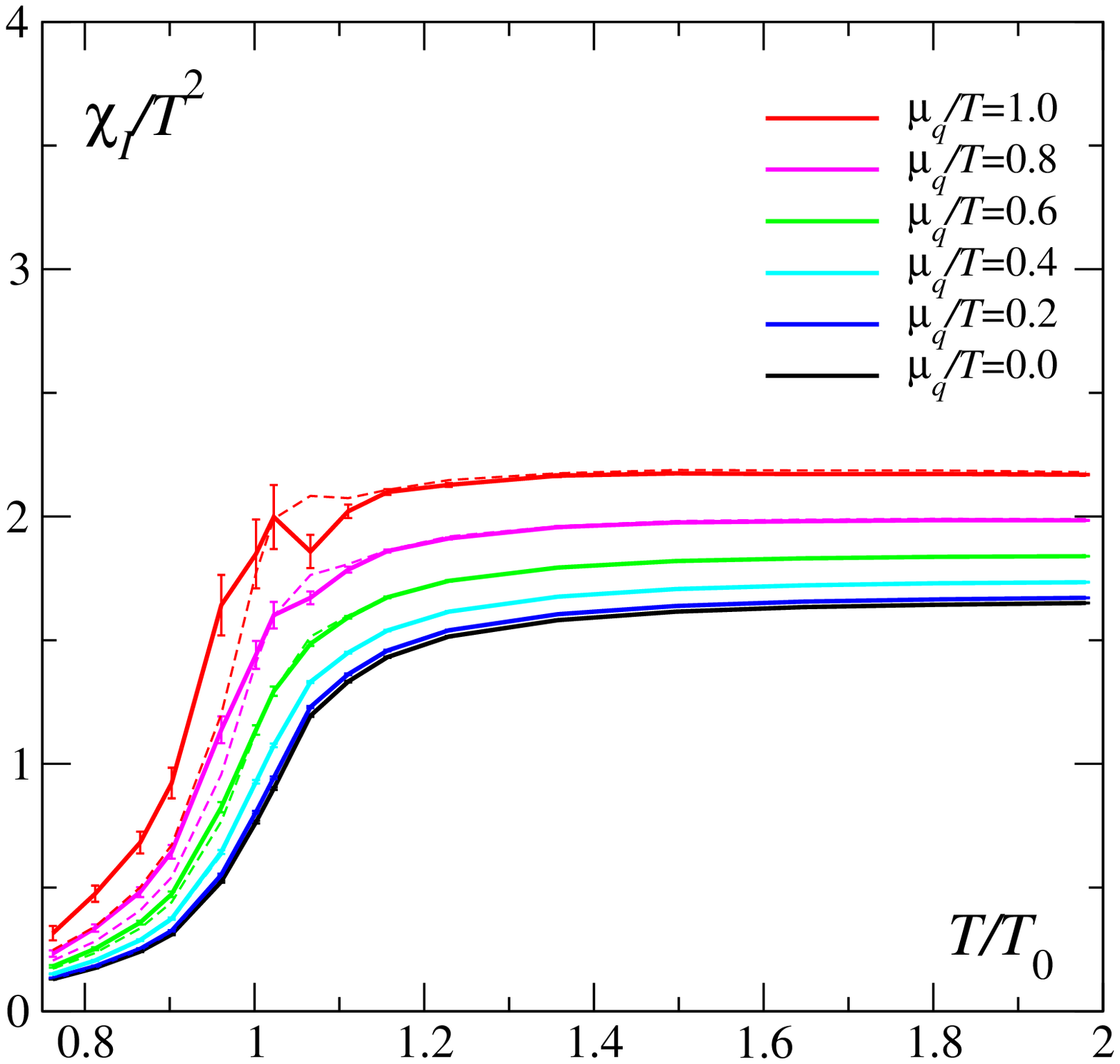}
\vskip -0.8cm
\caption{The quark number susceptibility $\chi_q/T^2$ (left) and 
isospin susceptibility $\chi_I/T^2$ (right) for various $\mu_q/T$.
$T_0$ is $T_c$ at $\mu_q=0$.}
\end{center}
\vskip -0.6cm
\label{fig:nsus}
\end{figure}

\section{Susceptibilities at non-zero $\mu_q$} 

Next, we calculate quark number and isospin susceptibilities 
in a range of $0 \leq \mu_q/T \leq 1$. 
The data connected by solid lines in Fig.~\ref{fig:nsus} are obtained by 
$\chi_q/T^2 = 2c_2 + 12c_4(\mu_q/T)^2 + 30c_6(\mu_q/T)^4$ 
and the corresponding equation for $\chi_I$. 
Dashed lines are the results from $O(\mu_q^2)$ expansion. 
Since the statistical error of $c_6$ is still large near $T_c$, 
the location and height of the peak are less accurately determined, 
and also the error from the truncation of higher order terms of 
$\mu_q/T$ seems to be visible for large $\mu_q/T$. 
However, as seen in Fig.~\ref{fig:c2c4c6}, 
$c_6$ changes its sign at $T_c$. 
This means the peak position of $\chi_q$ moves left, 
which is corresponding to the change of $T_c$ as a function of $\mu_q$. 
$T_c(\mu_q/T=1)/T_c(\mu_q/T=0)$ in \cite{tcmu} is about $0.93$. 
Moreover, $\chi_q$ increases with higher orders of the expansion 
for $T \le T_c$ which confirms the existence of a peak. 
This suggests the presence of a critical endpoint in the $(T,\mu_q)$ plane.
On the other hand, $\chi_I$ in Fig.~\ref{fig:nsus} does not show any 
singular behaviour. 
This is consistent with the sigma model prediction that only isosinglet 
degrees of freedom become massless at the critical endpoint.

\begin{figure}[tb]
\begin{center}
\includegraphics[angle=270,width=7.7cm]{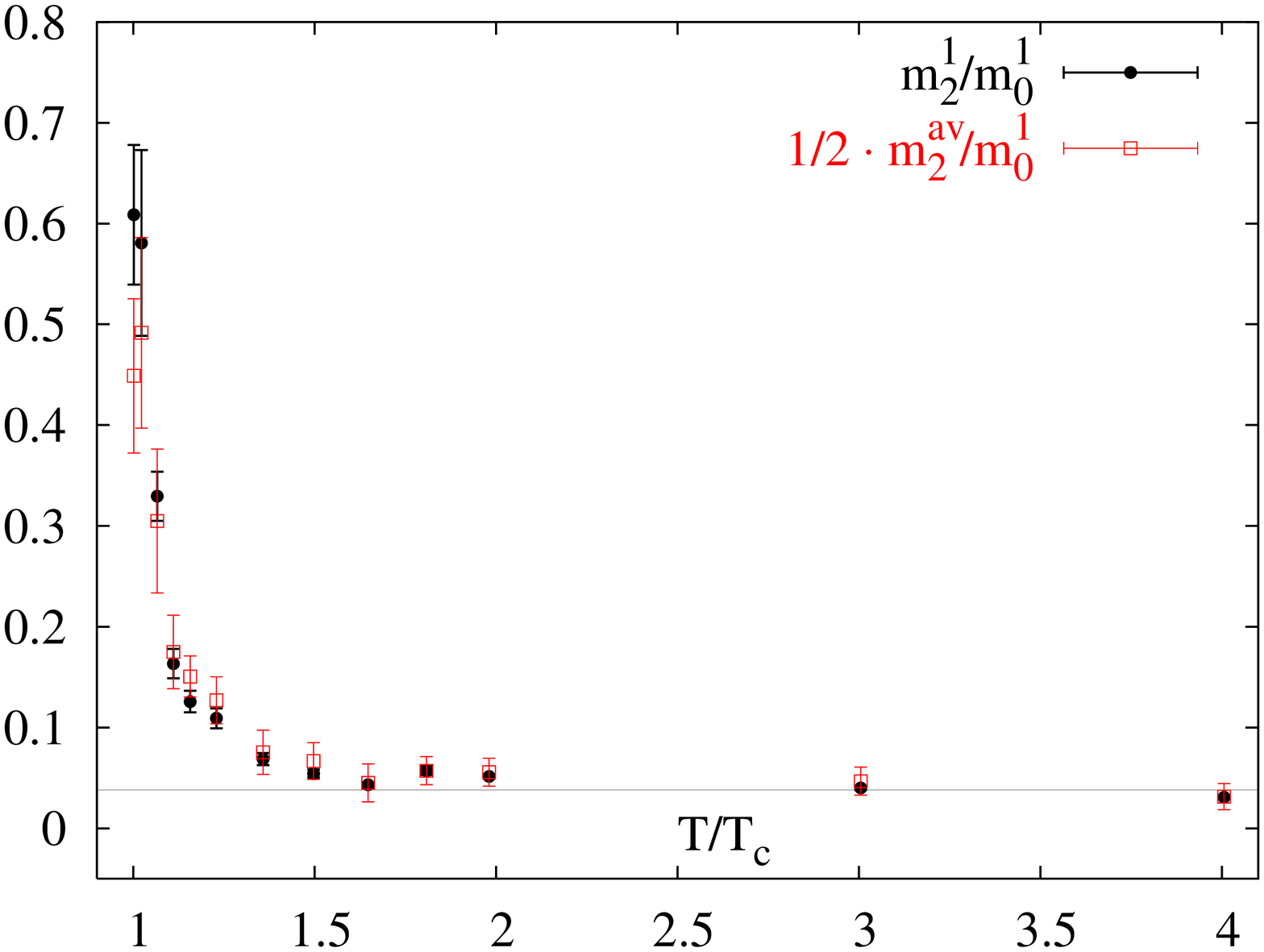} 
\includegraphics[angle=270,width=7.7cm]{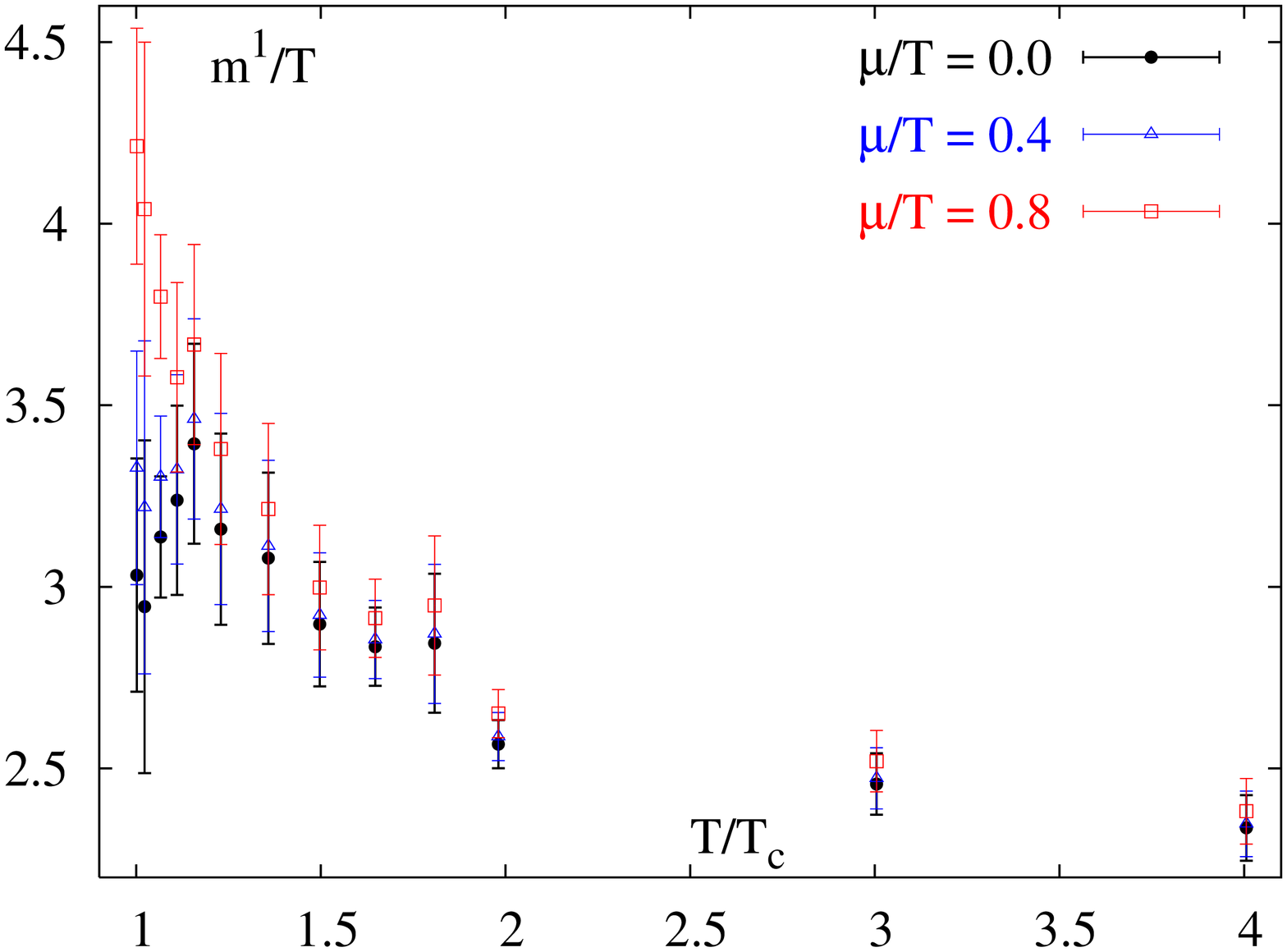} 
\vskip 0.0cm \hskip 0.5cm (a) \hskip 7.2cm (b)
\vskip -1.0cm
\caption{(a) The ratios $m_2^1/m_0^1$ and $m_2^{\rm av}/m_0^1$ as functions 
of $T/T_c$ and (b) the singlet screening mass at non-zero $\mu_q$.}
\end{center}
\vskip -0.8cm
\label{fig:m2m0}
\end{figure}

\section{Screening mass at non-zero $\mu_q$}

Finally we want to discuss the free energy of a static quark anti-quark 
pair at finite temperature and density. 
We extract the singlet free energy $F_{Q \bar{Q}}^1$ and 
colour averaged free energy $F_{Q \bar{Q}}^{\rm av}$ by 
the Polyakov loop correlation functions. For $T > T_c$,
the free energy is expected to be exponentially screened at large distances, 
$\Delta F^{\rm av,1}_{Q\bar{Q}}(r,T,\mu_q) \equiv 
F^{\rm av,1}_{Q\bar{Q}}(\infty,T,\mu_q) - F^{\rm av,1}_{Q\bar{Q}}(r,T,\mu_q) 
\sim e^{-m^{\rm av,1}(T,\mu_q) r}$. 
We calculate the Taylor expansion coefficients of the Debye screening mass 
$m^{\rm av,1}$ in terms of $\mu_q/T$ at $\mu_q=0$, 
$m^{\rm av,1} \equiv \sum_{n=0}^{\infty} m_n^{\rm av,1} (\mu_q/T)^n $.
We plot the data of $m_2^1/m_0^1$ and $m_2^{\rm av}/m_0^1$ in 
Fig.~\ref{fig:m2m0}(a). We find $m_2^1=(1/2)m_2^{\rm av}$. 
This is expected from perturbation theory, which suggests that 
the leading order contribution to the colour singlet free energy is 
given by one gluon exchange while the colour averaged 
free energy is dominated by two gluon exchange. 
Moreover the perturbative Debye screening mass $m_D$ is given by 
$ m_D(T, \mu_q) = m_{D,0}(T) \sqrt{1+3N_f/[(2N_c+N_f) \pi^2] (\mu_q/T)^2} $, 
with $m_{D,0}(T)=g(T)T \sqrt{(2N_c+N_f)/6}$. 
Here $N_c$ and $N_f$ are the number of colour and flavour respectively.
The solid line in Fig.~\ref{fig:m2m0}(a) is the perturbative prediction 
for $m_2^1/m_0^1$. 
The ratio $m_2^1/m_0^1$ is found to be consistent with perturbation 
theory for $T \geq 2T_c$. 
In Fig.~\ref{fig:m2m0}(b) we show the $\mu_q$-dependence of the singlet
screening mass $m^1(\mu_q,T)/T$ for a small values of $\mu_q/T$, 
where only contributions from $m_0^1$ and $m_2^1$ are included. 
Further details of this study are given in \cite{dsm}.

\end{document}